\documentstyle[12pt]{article}
\def\Journal#1#2#3#4{{#1} {\bf #2}, #3 (#4)}

\def\PLB{{\em Phys. Lett.}  B}
\def\PRL{\em Phys. Rev. Lett.} 
\def\PRD{{\em Phys. Rev.} D}
\def\NPA{{\em Nucl. Phys.} A}
\def\NPB{{\em Nucl. Phys.} B}
\def\PRC{{\em Phys. Rev.} C}
\def\JPG{{\em J. Phys.} G}
\def\CPL{\em Chin. Phys. Lett.}
\def\EPC{{\em Euro.Phys.J.} C}
\def\PTP{{\em Prog.Theor.Phys.} Suppl.}
\def\CTP{{\em Communi.Theor.Phys.}}
\def\HEPANP{{\em High Energy Phys. and Nucl. Phys.}}

\begin{document}

\begin{titlepage}

\vspace{1cm}

\centerline{\Large \bf Dibaryon Systems in Chiral $SU(3)$ }
\centerline{\Large \bf Quark Model\footnote{This work was
partly supported by the National
Natural Science Foundation of China (NSFC)
and the Chinese Academy of Sciences}}
\vspace{1cm}

\def\baselinestretch{0.5}

\centerline{\bf {Q.B.Li$^{a}$, P.N.Shen$^{b,a,c,d}$, Z.Y.Zhang$^{a}$,
Y.W.Yu$^{a}$}}
\vspace{0.5cm}

{\small
{
\flushleft{\bf  $~~~a.$ Institute of High Energy Physics, Academia Sinica,
 P.O.Box 918(4),}
\flushleft{\bf  $~~~~~~$Beijing 100039, China}

\def\baselinestretch{0.5}

%\vspace{8pt}
\flushleft{\bf  $~~~b.$ China Center of Advanced Science and Technology
 (World}
\flushleft{\bf  $~~~~~~$ Laboratory), P.O.Box 8730, Beijing 100080, China}

\def\baselinestretch{0.5}

%\vspace{8pt}
\flushleft{\bf  $~~~c.$ Institute of Theoretical Physics, Academia Sinica,}
\flushleft{\bf  $~~~~~~$ P.O.Box 2735, Beijing 100080, China}

\def\baselinestretch{0.5}

%\vspace{8pt}
\flushleft{\bf  $~~~d.$ Center of Theoretical Nuclear Physics, National Lab
of Heavy }
\flushleft{\bf  $~~~~~~$Ion Accelerator, Lanzhou 730000, China}

}}

\vspace{1.5cm}

\def\baselinestretch{1.5}

\centerline{\bf Abstract}

The possible candidates of $S-$wave dibaryons with various strange 
numbers are studied
under the chiral $SU(3)$ quark model. It is shown that there are three
types of baryon-baryon bound states. The states of the first type 
are called deuteron-like states. If chiral fields can provide enough
attraction between interacting baryons, these systems, such as 
$[\Xi\Omega - \Xi^{*}\Omega]_{(1,1/2)}$,
$[\Xi\Xi]_{(0,1)}$, $[N \Omega]_{(2,1/2)}$ would be weakly bound.
The states of the second type such as $[\Sigma^{*} \Delta]_{(0,5/2)}$,
$[\Sigma^{*} \Delta]_{(3,1/2)}$, $[\Delta\Delta]_{(0,3)}$ and $[\Delta\Delta]_{(3,0)}$ 
are named as $\Delta\Delta$-like states. Due to the highly 
symmetric character in orbital space, these systems could be relatively 
deeply bound, but the strong decay modes of composed 
baryons cause the widths of the states much broader. The states of 
the third type are entitled as $\Omega\Omega$-like states. Due to the same 
symmetry character shown in the systems of the second type and the only weak decay 
mode of composed baryons, for instance in $[\Omega\Omega]_{(0,0)}$, or at most one 
strong decay mode of composed baryons, for example in
$[\Xi^{*}\Omega]_{(0,1/2)}$, these states are deeply bound states with narrow
widths. The states of latter two types are most interesting new dibaryon
states and  should be carefully investigated both theoretically and
experimentally. \end{titlepage}

\baselineskip 18pt

\noindent
{\bf 1. Introduction.}

Dibaryon as a six-quark system has shown its special place in the 
investigation of medium-energy physics since Jaffe published his prediction 
of $H$ particle in 1977 \cite{ja}. As is commonly believed, to study a
quark system, the Quantum Chromodynamics ($QCD$), which governs the strong
interaction among quarks and gluons, should be employed as an underline theory although
its non-perturbative behavior is still not quite clear and cannot 
exactly be solved up to now. Jaffe studied the color-magnetic interaction 
($CMI$) of the one-gluon-exchange $(OGE)$ potential in the multi-quark 
system and found that $CMI$ shows attractive character in the $H$ particle 
case. This character compels six quarks staying in a smaller volume,
say less than $0.85fm$ in radius. Thus dibaryon study could provide more information 
about the short-range behavior of $QCD$, and the existence of dibaryon
can directly supply the evidence of the quark-gluon degrees of freedom
in hadrons and hadronic systems. 
However, the reason for forming the baryon-baryon bound state presents great 
complexity. The nonperturbative $QCD$ ($NPQCD$) effect may seriously 
affect the properties of the dibaryon due to its finite size. The $CMI$
and the interaction describing the action 
from physical vacuum should be co-responsible for the binding energy 
of the system. The symmetry structure of the system may also play
an important role there. In a word, the character of CMI is no longer dominant. 
Sometimes, meson clouds may provide predominant effects. Exploring dibaryon
may enable us to investigate the short- and medium-range $NPQCD$ effects and 
to find out a practical way to properly treat them.   

\vspace{0.3cm}

In order to reliably study dibaryon, a model that can describe most of short- and medium-range $NPQCD$ effects should be employed. In other word, the 
model should have predictive power. It should at least contain two-fold requirements: By using this model the ground state properties of baryons should well 
be fitted, and the experimental data of the nucleon-nucleon ($N-N$) scattering and, especially, the  empirical data of the nucleon-hyperon ($N-Y$) scattering 
and reaction can  reasonably be reproduced in the dynamical calculation. 
When extending 
this model to dibaryon investigation, no additional parameters are required.
There are lots of models such as MIT bag model
\cite{mit}, the constituent quark models of various kinds 
\cite{clu,zys,ysz}, Skyrme model \cite{sky}, etc. Among them, the chiral 
$SU(3)$ quark model \cite{zys,ysz} is one of the models which satisfy 
requested conditions. In
terms of this model, investigating and further systematically analyzing 
possible bound six-quark systems become significant and essential. 

\vspace{0.3cm}

Since 70's, dibaryon has been intensively studied.
The most interesting dibaryons have been studied are the following:
$H$ particle has
been theoretically and experimentally investigated for years. The
theoretically predicted mass is in a large range \cite{ja,clu,ya,szy,sk,oy,mul}, say
from $2GeV$ to $2.4GeV$. The most believed theoretical prediction is around
the $\Lambda\Lambda$ threshold, namely around $2.232GeV$ \cite{ya,szy}.
However, this particle still has not been found in the experiment
yet. The most recent data showed that the lower limit of
the $H$ particle mass is about $2.22GeV$ \cite{im}. 
Except the $H$ particle, possible bound baryon-baryon systems in the 
non-strangeness sector were also investigated. $d^*$ is one of them.
There were number of theoretical predictions by using various models, such as
the non-relativistic boson-exchange model \cite{ka}, the quark cluster model
\cite{ya}, the quark-delocation model \cite{wan},
the chiral $SU(3)$ quark model \cite{yzys} and etc.. The predicted
masses also spread in a wide range. All the predictions are below 
the threshold of the $\Delta\Delta$ channel
of $2.464GeV$, and most of them are above the threshold of the
strong decay channel, $N N \pi\pi$, of $2.154GeV$. $d'(J^{P}=0^{-},T=0)$
is another interesting particle. In the experiments of double-charge-exchange 
reactions, 
it was found that when the energy of the incident pion is $50MeV$, there exists
a resonance with the mass of $2.065GeV$ and the width of $0.5MeV$
in the processes with a variety of targets \cite{bil}. To explain that 
phenomenon, one proposed $d'$. Although there were many theoretical
attempts \cite{wag}, the theoretical result is still away from the 
expected value. Whether this phenomenon indicates $d'$ is still under
discussion. Up to now, these three interesting candidates of dibaryons
are still not found or confirmed by experiments. It seems that one should go beyond
these candidates and should search the possible candidates in a 
wider region, especially the systems with multi-strangeness, in terms 
of a more reliable model such as chiral $SU(3)$ quark model. According to 
this idea, Yu et al. analyzed the six-quark system with a simple
six-quark cluster configuration \cite{yzy}.
Later, by employing the chiral $SU(3)$ quark model, Zhang, Yu et al. 
studied $\Omega\Omega(S=0,T=0)$ and $\Xi\Omega(S=1,T=1/2)$ \cite{yzy,yzy1}, 
and Li and Shen explored $\Xi^{*}\Omega(S=0,T=1/2)$ and 
$\Xi\Omega-\Xi^{*}\Omega(S=1,T=1/2)$ \cite{ls}. In this paper, we would
present a systematic study of possible candidates of $S-$wave baryon-baryon bound states
in this model.

\vspace{0.3cm}

The paper is arranged in the following way: The chiral $SU(3)$ quark model
is briefly introduced in Sect.2. In Sect.3, the results calculated 
by this model are given, and the symmetry characters of the system concerned
are discussed. The effects of chiral-quark field induced interactions on
the binding behaviors of systems are detailed analyzed in Sect.4. In Sect.5,
the model parameter dependence of the predicted binding energy is further 
studied. Finally, in Sect.6, the concluding remark is drawn.

\vspace{0.5cm}

\noindent
{\bf 2. Brief introduction of chiral $SU(3)$ quark model.}

\vspace{0.3cm}

Following Georgi's idea \cite{geo}, the quark-chiral $SU(3)$ field 
interaction can be written as 
\begin{eqnarray}
{\cal L}_{I} & = &-g_{ch} (\bar{\psi}_{L} \Sigma \psi_{R} 
- \bar{\psi}_{R} \Sigma^{+}\psi_{L}) ~~,
\end{eqnarray}
with $g_{ch}$ being the quark-chiral field coupling constant, $\psi_{L}$ and
$\psi_{R}$ being the quark-left and right spinors, respectively, and 
\begin{eqnarray}
\Sigma = exp{[i \pi_{a} \lambda_{a}/f]}, ~~ a = 1,2, ...8~~.
\end{eqnarray}
where $\pi_{a}$ is the Goldstone boson field and $\lambda_{a}$ the Gell Mann
matrix of the flavor SU(3) group. Generalizing the linear realization of $\Sigma$
in the $SU(2)$ case to the $SU(3)$ case, one obtains
\begin{eqnarray}
\Sigma = \sum^{8}_{a=0} \sigma_{a} \lambda_{a} - i \sum^{8}_{a=0} \pi_a
\lambda_a ~~,
\end{eqnarray}
and the interaction Lagrangian 
\begin{eqnarray}
{\cal L}_{I}  =  -g_{ch} \bar{\psi} \left ( \sum^{8}_{a=0} \sigma_a \lambda_a +
i \sum^{8}_{a=0} \pi_a \lambda_a \gamma_5 \right ) \psi ~~,
\end{eqnarray}
where $\lambda_{0}$ is a unitary matrix, $\sigma_{0},..,\sigma_{8}$ the
scalar nonet fields and $\pi_{0},..,\pi_{8}$ the pseudo-scalar nonet fields.
Clearly, ${\cal L}_{I}$ is invariant under the infinitesimal chiral
$SU(3)_{L} \times SU(3)_{R}$ transformation. Consequently, one can write 
the interactive Hamiltonian as
\begin{eqnarray}
H_{ch} = g_{ch} F(q^{2}) \bar{\psi} \left ( \sum^{8}_{a=0} \sigma_a \lambda_a
+ i \sum^{8}_{a=0} \pi_a \lambda_a \gamma_5 \right ) \psi~~. 
\end{eqnarray}
Here we have inserted a form factor $F(q^{2})$ to describe the chiral field
structure \cite{obu}. As usual, $F(q^{2})$ is taken as
\begin{eqnarray}
F(q^{2}) = \left ( \frac{\Lambda^{2}_{CSB}}{\Lambda^{2}_{CSB}+q^{2}}
\right )^{1/2} ~~, 
\end{eqnarray}
and the cut-off mass $\Lambda_{CSB}$ indicates the chiral symmetry breaking scale \cite{obu}.

\vspace{0.3cm}

Then, the $SU(3)$ chiral-field-induced quark-quark potentials can be derived 
in the following :
\begin{eqnarray}
 V_{\sigma_a} (\vec{r}_{ij}) & = &-C(g_{ch}, m_{\sigma_a}, \Lambda_{CSB}) 
X_{1}(m_{\sigma_a}, \Lambda_{CSB}, r_{ij}) (\lambda_{a}(i) 
\lambda_a (j) ) \nonumber\\
  & + & V^{\vec{l} \cdot \vec{s}}_{\sigma_a} (\vec{r}_{ij}),\\
   V_{\pi_a}(\vec{r}_{ij}) & = & C(g_{ch}, m_{\pi_a}, \Lambda_{CSB}) 
\frac{m_{\pi_a}^{2}}{12m_{qi}m_{qj}} [ X_{2}(m_{\pi_a}, \Lambda_{CSB},
r_{ij}) (\vec{\sigma}_{i} \cdot \vec{\sigma}_{j})  \nonumber\\
  & + & \left ( H(m_{\pi_a} r_{ij}) - (\frac{\Lambda_{CSB}}{m_{\pi_a}} )^{3}
H(\Lambda_{CSB} r_{ij} ) \right ) S_{ij} ]  (\lambda_{a}(i)
\lambda_{a}(j) ) ~~,
\end{eqnarray}
with
\begin{eqnarray}
  V^{\vec{l} \cdot \vec{s}}_{\sigma_{a}} (\vec{r}_{ij}) & = & -C(g_{ch},
        m_{\sigma_{a}}, \Lambda_{CSB})
        \frac{m^{2}_{\sigma_{a}}}{4m_{qi} m_{qj}} 
        [ G (m_{\sigma_{a}} r_{ij} )  \nonumber\\
  & - & (\frac{\Lambda_{CSB}}{m_{\sigma_{a}}} )^{3}  
        G(\Lambda_{CSB} r_{ij})] 
        (\vec{L} \cdot (\vec{\sigma}_{i} + \vec{\sigma}_{j} ))
        (\lambda_{a}(i)\lambda_{a}(j) ) ~~,\\
 C(g_{ch} , m, \Lambda)& = & \frac{g^{2}_{ch}}{4\pi}
          \frac{\Lambda^{2}}{\Lambda^{2} - m^{2} } m ~~,\\
 S_{ij}& = & 3(\vec{\sigma}_{i} \cdot \hat{r}) 
        (\vec{\sigma}_{j} \cdot \hat{r})
      -(\vec{\sigma}_{i} \cdot \vec{\sigma}_{j}) ~~,
\end{eqnarray}
and $m_{\sigma_a}$ being the mass of the scalar meson and $m_{\pi_a}$
the mass of the pseudo-scalar meson. 
The explicit forms of functions $X_1,~X_2,~H$ and $G$ can be found in 
\cite{zys}. 

\vspace{0.3cm}

As mentioned in \cite{zys}, the interactions induced by chiral
fields describe the NPQCD effect in the low-momentum
medium-distance range, which is very important in explaining 
the short- and  medium-range forces between two baryons. To study the 
baryon structure and
baryon-baryon dynamics, one still needs an effective
one-gluon-exchange interaction $V^{OGE}_{ij}$ which dominates 
the short-range perturbative QCD behavior and a confinement potential 
$V^{conf}_{ij}$ which provides the NPQCD effect in the 
long distance and confines three quarks to a baryon. Then, the total 
Hamiltonian of the six-quark system can be written as
\begin{eqnarray}
H = \sum_{i} ~~T_{i} - T_{G} + \sum_{i<j} V_{ij} ~~,
\end{eqnarray}
with
\begin{eqnarray}
V_{ij} = V^{OGE}_{ij} + V^{conf}_{ij} + V^{ch}_{ij} ~~,
\end{eqnarray}
where
\begin{eqnarray}
 & V^{OGE}_{ij} & = \frac{1}{4} g_{i}g_{j} (\lambda^{c}_{i} \cdot
       \lambda^{c}_{j})
       \{ \frac{1}{r_{ij}} - \frac{\pi}{2} \delta(\vec{r}_{ij})
       ( \frac{1}{m^{2}_{qi}} + \frac{1}{m^{2}_{qj}} \nonumber \\
   & & + \frac{4}{3} \frac{1}{m_{qi}m_{qj}} (\vec{\sigma_{i}}
        \cdot \vec{\sigma_{j}}) )
        -\frac{1}{4m_{qi}m_{qj}r^{3}_{ij}} S_{ij} \} +
        V^{\vec{\ell}\cdot\vec{s}}_{OGE}~~,\\
 & V^{\vec{\ell} \cdot \vec{s}}_{OGE}&  = -\frac{1}{16} g_{i}g_{j}
         (\lambda^{c}_{i} \cdot \lambda^{c}_{j})
         \frac{3}{m_{q_i}m_{q_j}} \frac{1}{r^{3}_{ij}} \vec{L} 
         \cdot (\vec{\sigma}_{i}
         + \vec{\sigma}_{j} ) ~~,\\
 & V^{conf}_{ij} & = -a^{c}_{ij} (\lambda^{c}_{i} \cdot \lambda^{c}_{j})
           r^{2}_{ij} -a^{c0}_{ij} (\lambda^{c}_{i} \cdot 
           \lambda^{c}_{j})~~,
\end{eqnarray}
and
\begin{eqnarray}
V^{ch}_{ij} = \sum^{8}_{a=0} V_{\sigma_a} (\vec{r}_{ij}) + \sum^{8}_{a=0}
   V_{\pi_a} (\vec{r}_{ij}) ~~.
\end{eqnarray}

\vspace{0.3cm}

The model parameter should be fixed before calculation. 
The coupling constant $g_{ch}$ is fixed by
\begin{eqnarray}
\frac{g^{2}_{ch}}{4\pi} = ( \frac{3}{5} )^{2} \frac{g^{2}_{NN\pi}}{4\pi}
\frac{m^{2}_{q}}{M^{2}_{N}} ~~, 
\end{eqnarray}
and $g^{2}_{NN\pi}/4\pi$ is taken to be the empirical value of about 14.
The masses of the pseudo-scalar
meson $m_{\pi}, m_{\eta}, m_{\eta'}$ and $m_{K}$ can be chosen
as the masses of the real $\pi$, $\eta$, $\eta^{\prime}$ and 
$K$, and the mass of the scalar meson $\sigma_{0}$ can be taken as 
$m_{\sigma_0} \simeq 625 ~MeV$,
according to the relation \cite{sca}
\begin{eqnarray}
m^{2}_{\sigma_0} = (2m_{q})^{2} + m^{2}_{\pi} ~~.
\end{eqnarray}
In our previous investigation \cite{zys}, it was found the $N-N$ and $N-Y$ 
scatterings are not sensitive to the masses of strange chiral fields. 
In order to reduce the numbers 
of adjustable parameters, 
$m_{\sigma_a}(a=1,..,8)$ are also taken to be the mass of $\eta'$.
The cut-off mass $\Lambda_{CSB}$ for various chiral fields is taken to be
\begin{eqnarray}
\Lambda_{CSB} = \left\{
\begin{array}{ll}
4.2 fm^{-1}, & \mbox{~~${\rm for} ~\pi,  K~  {\rm and}~  \sigma_{0}$} \\
5.0 fm^{-1}, & \mbox{~~${\rm for} ~\eta, \eta', \sigma', \kappa~
{\rm and}~  \epsilon.$}
\end{array}
\right.
\end{eqnarray}
This set of model parameters is called Model I (Set I) which was frequently
used in our pervious investigations \cite{zys,ysz,szy}. Because the systems 
studied in this paper mostly comprise strange quarks, the 
strange chiral clouds surrounding the baryons become influential. 
In order to see this effect,
we increase the masses and corresponding cut-masses of 
$\kappa$ and $\epsilon$ to the values of $1.4GeV$ and $1.5GeV$, respectively, 
which are close to the masses of real mesons with the same quantum numbers
\cite{pdg}. This set of parameters is called Model I (Set II).
When the values of $m_{u}, m_{s}, b_{u}, g_{ch}, m_{\pi_a}, m_{\sigma_a}$ and
$\Lambda_{CSB}$ are fixed, the one gluon exchange coupling constants
$g_{u}$ and $g_{s}$ can be determined by  mass splittings between
$\Delta$ and  $N$, and $\Sigma$ and  $\Lambda$, respectively, 
the confinement
strengths $a^{c}_{uu}$, $a^{c}_{us}$ and $a^{c}_{ss}$ are fixed by the
stability conditions of $N$, $\Lambda$ and $\Xi$, respectively, and the
zero point energies  $a^{c0}_{uu}$, $a^{c0}_{us}$ and $a^{c0}_{ss}$ 
are fixed by the masses of $N$, $\Lambda$ and $\overline{\Sigma + \Omega}$,
respectively. 

As is mentioned above, to predict the dibaryon structure, the model
should be able to reproduce the data of the N-N and Y-N scatterings
reasonably. The detailed comparison of the
theoretical scattering results and the empirical data can be found in 
\cite{zys}. Here, we only show a typical plot, the 
cross section of the $\Lambda-p$ process in Fig.1.
In this figure, the solid and dashed curves represent the results with Sets I 
and II, respectively.
It is shown that both curves are consistent with the
experimental data. After confirming the model, we use the same set of 
parameters to study the dibaryon system.

%\vspace{0.3cm}

%{\scriptsize {\bf Fig.1. Differential cross section of $Lambda-p$ process}. 
%The solid and dashed curves denote the results from the set I and Set II,
%Respectively.}

%\vspace{0.3cm}

All the model parameters in Sets I and II are tabulated in Table 1, respectively.

\vspace{0.3cm}

{\flushleft {\bf Table 1. Model parameters}}
\begin{center}
\tabcolsep 3pt
\begin{tabular}{|c||c|c|}
\hline
{$~$} & Set I & Set II\\ \hline
$m_{u}(MeV)$            & 313   &  313\\
$m_{s}(MeV)$            & 470   &  470\\
$b_{u}(fm)$             & 0.505 &  0.505\\
$g_{u}$                 & 0.936 &  0.936\\
$g_{s}$                 & 0.924 &  0.781\\
$a^{c}_{uu}(Mev/fm^2)/ a^{c0}_{uu}(Mev)$  & 54.3/-47.7  &  55.0/-48.9\\
$a^{c}_{us}(Mev/fm^2)/ a^{c0}_{us}(Mev)$  & 65.8/-41.7 &  66.5/-50.6\\
$a^{c}_{ss}(Mev/fm^2)/ a^{c0}_{ss}(Mev)$  & 103.0/-50.6 &  115.4/ -73.7\\
$m_{\pi}(fm^{-1})/\Lambda_{\pi}(fm^{-1})$  & 0.7/4.2 & 0.7/4.2\\
$m_{K}(fm^{-1})/\Lambda_{K}(fm^{-1})$  & 2.51/4.2 & 2.51/4.2\\
$m_{\eta}(fm^{-1})/\Lambda_{\eta}(fm^{-1})$  & 2.78/5.0 & 2.78/5.0\\
$m_{\eta'}(fm^{-1})/\Lambda_{\eta'}(fm^{-1})$ & 4.85/5.0 & 4.85/5.0\\
$m_{\sigma_0}(fm^{-1})/\Lambda_{\sigma_0}(fm^{-1})$  & 3.17/4.2 & 3.17/4.2\\
$m_{\sigma'}(fm^{-1})/\Lambda_{\sigma'}(fm^{-1})$  & 4.85/5.0 & 4.85/5.0\\
$m_{\kappa}(fm^{-1})/\Lambda_{\kappa}(fm^{-1})$  & 4.85/5.0 & 7.09/7.6\\
$m_{\epsilon}(fm^{-1})/\Lambda_{\epsilon}(fm^{-1})$  & 4.85/5.0 & 7.09/7.6\\
\hline
\end{tabular}
\end{center}

\vspace{0.5cm}

The binding energy of the baryon-baryon system is dynamically 
solved by using the
Resonating Group Method ($RGM$). In this method, the trial wave function of 
the six-quark system can be written as
\begin{eqnarray}
\Psi={\cal A}[ \hat{\phi_A}(\xi_{1},\xi_{2}) \hat{\phi_B}(\xi_{4},\xi_{5})
 \chi_{rel} (\vec{R}) \chi_{CM}(\vec{R}_{CM})]_{ST},
\end{eqnarray}
where $\phi_{A(B)}$ denotes the antisymmetrized wave function of the
baryon cluster A(B), $\chi_{rel} (\vec{R})$ the trial wave function of 
the relative motion
between interacting clusters $A$ and $B$, $\chi_{CM}(\vec{R}_{CM})$ the
wave function of the motion of the total center of mass, and $\xi_{i}$ the
Jacobi coordinate with $i=1$ and $2$ for cluster $A$ and $i=4$ and $5$ for
cluster $B$, respectively.  The symbol 
${\cal A}$ describes the operation of the antisymmetrization between quarks
in two interacting clusters. This operator can be read as
\begin{eqnarray*}
{\cal A} = N \sum_{P} \epsilon_{P} P,
\end{eqnarray*}
where $P$ is an operator which permutes quarks of cluster $A$ and quarks 
of cluster $B$,
$\epsilon_P =1(-1)$ when $P$ is an even(odd) permutation and $N$ is the 
normalization factor. Considering the permutation symmetry, $\cal A$ 
can also be written as
\begin{eqnarray}
{\cal A} &=& N'(1 - \sum_{\stackrel{i\in A}{j \in B}}P_{ij}^{osfc}), 
\end{eqnarray}
where $P_{ij}^{osfc}$ denotes the permutation operation 
carried out between the $i-th$ and $j-th$ quarks in the $orbit$, 
$spin$, $flavor$ and $color$ spaces, simultaneously, and again 
$N'$ represents the normalization constant. Calculating the expectation
value of the Hamiltonian operator on the trial wave function in which
the unknown $\chi_{rel}(\vec{R})$ is expanded in terms of well-known 
bases, one deduces a secular equation
\begin{eqnarray}
\sum_{j=1}^{n}(H_{ij} - E N_{ij})c_{j}=0, \  \  \  \  (i=1,2,\cdots,n),
\end{eqnarray}
with $H_{ij}$ and $N_{ij}$ being the Hamiltonian and normalization matrix
elements, respectively, $E$ the eigenvalue and $c_{i}$ the corresponding 
eigenfunction, namely the expansion coefficients of $\chi_{rel}(\vec{R})$.
Solving this equation for $c_{j}$, one obtains the binding energy and the
corresponding wave function of the six-quark system. The detailed method can 
be found in our previous paper \cite{ls,szy}. 

\vspace{0.5cm}

\noindent
{\bf 3. Symmetry character discussion.}

\vspace{0.3cm}

As a comprehensive survey, there are two crucial physical factors 
which resolve whether a two-baryon system is bound.
One is the symmetry property of the system,
namely the characteristics of quark exchanges between baryons,
and the other is the interaction including both the direct and exchange 
components between quarks, and eventually between baryons. 
In this section, we analyze the symmetry property of the system 
according to the character of the matrix element $P_{36}^{sfc}$, 
where the superscript $sfc$ denotes the $P$ operator acts within the 
$spin-flavor-color$ space only and the subscript $36$ represents the
exchange operation is between the 3-rd and 6-th quarks. Then, in section 4,
we discuss the effects of various chiral-quark field interactions by
employing several models.

\vspace{0.3cm}

As is pointed out in Ref.[21], the matrix element $\langle {\cal A}^{sfc} 
\rangle$ is an important measure of the action of the Pauli principle 
in the two-baryon state. This measure specifies the symmetry character of 
the state of the system. According to the symmetry characters of systems, 
namely the mentioned matrix elements, the two-baryon systems concerned can 
generally be divided into  
three classes. In the first class, $9 \langle P_{36}^{sfc} \rangle\sim 1$,
namely, $\langle {\cal A}^{sfc} \rangle\sim 0$. The Pauli blocking effect
between interacting baryons are incrediblely serious so that the two-baryon
$S-$wave state with $[6]_r$, the $[6]$ symmetry
in the orbital space, is almost a forbidden state. Namely, it is very hard to form a bound state.
The state in the second class has the property of  
$9 \langle P_{36}^{sfc} \rangle\sim 0$, or 
$\langle {\cal A}^{sfc} \rangle\sim 1$. In this class, the Pauli blocking 
effect between interacting baryons are very small so that the exchange
effect between  quarks which are located in different baryons, respectively,
becomes negligible and these two baryons are relatively
independent with each other. As a result, the meson-exchange effect may
play a dominant role in binding. If the inter-baryon interaction 
shows an attractive feature with a large enough strength, the system would be bound. 
This kind of system may also be deduced in terms of a model in the baryon-meson
degrees of freedom. The state in the third class possesses a feature of
$9 \langle P_{36}^{sfc} \rangle\sim -1$, or 
$\langle {\cal A}^{sfc} \rangle\sim 2$. The inter-baryon quark-exchange 
feature of this kind would be enormously beneficial to form a state with 
the $[6]_r$ symmetry \cite{zyc}. If the inter-baryon interaction demonstrates
the attractive character with certain strength, it is possible to form
not only a bound state, but also a dibaryon with a relative smaller
size in radius. 
We present the resultant binding energies and the
corresponding root-mean-square radii ($RMS$) of the $S-$states which
have various strange number ${\cal{S}}$ and belong to the second and third classes
in the following subsections.

\vspace{0.3cm}

\noindent
{\bf 3.1. The systems in which the expectation values of $P^{sfc}_{36}$ operator
are close to zero.}

\vspace{0.3cm}

In the deuteron case, $\langle P^{sfc}_{36} \rangle=-1/81$,
it is a typical case of the second class.
We collect some systems which have the same symmetry characteristics as
deuteron in this subsection. The binding energies, $E_b$, and corresponding 
$RMS$, ${\cal R}$, of these systems are tabulated in Table 2.  

\vspace{0.3cm}

\begin{scriptsize}
{\flushleft{\bf {Table 2. Binding energy, $E_b$, and corresponding $RMS$,  
${\cal R}$, for the systems with $\langle P^{sfc}_{36} \rangle \approx 0$}}.
The units for $E_b$ and RMS are in $MeV$ and $fm$, respectively.}

\begin{center}
\tabcolsep 3pt
\begin{tabular}{|c|c|c||c|c|c|}
\hline
${\cal{S}}$ & \multicolumn{2}{|c||}{System} & $\langle P^{sfc}_{36} \rangle$
         & Model I (Set I) & Model I (set II) \\
& \multicolumn{2}{|c||}{} & & $E_b~~//{\cal R}$ & $E_b~~//{\cal R}$ 
     \\ \hline
  0 & \multicolumn{2}{|c||}{$ NN(S=1,T=0)$} &  &  & \\
    & \multicolumn{2}{|c||}{deuteron} & $-1/81$ & 0.2~//1.63 & -0.7~//1.68 
         \\ \hline 
    & $N\Lambda$ & $ N \Lambda(S=0,T=1/2)$ & $0$ &  -6.6~//1.63 & -7.7~//1.69
         \\ \cline{3-6}
& $(S=0,T=1/2)$& $ N \Lambda-N \Sigma(S=0,T=1/2)$ &  & -5.7~//1.59 
     & -7.2~//1.66 \\ \cline{2-6}
 -1 & $N\Lambda$& $ N \Lambda(S=1,T=1/2)$ & $0$ & -7.7~//1.68 & -7.7~//1.68
         \\ \cline{3-6}
& $(S=1,T=1/2)$& $ N \Lambda-N \Sigma(S=1,T=1/2)$ &  & -6.6~//1.63  
     & -6.7~//1.64  \\ \cline{2-6}
    & \multicolumn{2}{|c||}{$ N \Sigma(S=0,T=3/2)$} & $-1/81$ & -5.1~//1.58 
         & -5.3~//1.59 \\ \hline
    & $\Lambda\Lambda$ &$\Lambda\Lambda(S=0,T=0)$ & $0$ & -4.8~//1.54 
         & -5.6~//1.58\\ \cline{3-6}  
 -2 & $(S=0,T=0)$ & $\Lambda\Lambda-N\Xi-\Sigma\Sigma(S=0,T=0)$ 
         &  &  &  \\  
    & & $H$ particle &    & -2.0~//1.15 
         & 8.2~//0.91 \\ \hline
    & \multicolumn{2}{|c||}{$\Lambda\Xi(S=1,T=1/2) $} & 4/81 & -8.1~//1.74 
         & -7.7~//1.72 \\ \cline{2-6}
 -3 & \multicolumn{2}{|c||}{$N\Omega(S=2,T=1/2)$} & 0 & 3.5~//1.18 
         & 12.7~//0.98\\ \cline{2-6}
    & \multicolumn{2}{|c||}{$\Delta\Omega (S=3,T=3/2)$} & 0 
         & 4.4~//1.15 & 14.2~//0.96\\  \hline
 -4 & \multicolumn{2}{|c||}{$\Xi\Xi(S=0,T=1)$} & $-1/81$ 
         & 4.1~//1.17  & 0.4~//1.30\\  \hline
 -5 & $\Xi\Omega$ & $\Xi\Omega(S=1,T=1/2)$ & $1/81$ 
         & 9.5~//1.02 & 4.2~//1.14\\ \cline{3-6}
    & $(S=1,T=1/2)$ & $\Xi\Omega-\Xi^*\Omega(S=1,T=1/2)$ & 
         & 32.9~//0.78 & 32.6~//0.77\\ \hline    
\end{tabular}
\end{center}
\end{scriptsize}
\vspace{0.5cm}

\noindent
The data in Table 2 shows that the deuteron is weakly bound in the Model I
(Set I) case, which indicates that the chiral $SU(3)$ quark model, in 
principle, can reasonably describe the structure of deuteron{\footnote {
Of course, it is easy to fine-tune the mass of $\sigma_0$ so that the best 
agreement between the calculated and experimental binding energy of deuteron
can be achieved. However, it is not necessary to give a very accurate mass
of predicted dibaryon, we would relinquish this adjustment.}}.
It is also seen that in the single $N\Lambda$ channel and the coupled 
$N\Lambda-N\Sigma$ channel with $S=0 (or 1)$ and $T=1/2$ cases and
the single $ N \Sigma$ channel with $S=0$ and $T=3/2$ case,
no matter which set of model parameters is employed, the systems are not
bound.  These are in agreement with experiments.
The $H$ particle is also not bound in Set I but weakly bound in Set II. The 
resultant mass of H is close
to the $\Lambda\Lambda$ threshold in both Set I
and Set II, and this feature is consistent with the recent finding in 
experiments \cite{im}. These results further convince ourselves that the
chiral $SU(3)$ quark model is reasonable and reliable in the
bound-state study.

\vspace{0.3cm}

It should be noted that $\langle P^{sfc}_{36} \rangle$ of these systems 
appoximately being zero means that the symmetry structure of this
kind makes the quark-exchange effect less important, and consequently,
the contribution from the kinetic energy term shows relatively repulsive nature
to the kinetic energy of the relative motion between  two well-separated 
interactive baryons (see Appendix), which
makes interacting baryons apart.  Therefore, very similar
to deuteron, whether the two-baryon system is bound depends on the feature
of the interaction between interacting baryons, especially 
that caused by chiral fields, namely the overall 
characteristics of the short- and medium-range $NPQCD$ effects, to a 
considerable extent. If the characteristics is attractive in nature,
the system would be bound like deuteron, and we call it as a deuteron-like 
system. If it shows weak
attraction or even repulsive feature, the system would not be bound
anymore. It is also noticed that the $OGE$ interaction provides repulsion 
in the
deuteron, $N \Lambda$, $N \Sigma$, $\Xi\Xi(S=0,T=1)$ and $\Xi\Omega(S=1,T=1/2)$
systems. To form a 
bound state, a strong enough attractive interaction by the meson
exchange must be requested. For instance, the $\pi$ exchange, especially
the tensor force, causes the weakly binding of deuteron and
the relative strong attraction from the coupling between the chiral fields
and quarks makes $\Xi\Xi(S=0,T=1)$ and $\Xi\Omega(S=1,T=1/2)$ bound.
Then in the $H$ particle case,
although $OGE$ interaction provides attraction, due to the repulsive effect
from the kinetic energy term, the only attraction from $OGE$ potential is 
not strong enough to cause binding. The attractive feature of the chiral 
field would be helpful to form a weakly bound $H$. But it is model parameter
dependent. In the $N \Omega(S=2,T=1/2)$ and $\Delta\Omega(S=3,T=3/2)$
cases, $OGE$ contributes nothing, the weakly bound behavior of these systems
fully depends on the attractive features of the chiral fields.
Moreover, it is noteworthy that very similar to the deuteron,
the $\Lambda\Lambda-N\Xi-\Sigma\Sigma(S=0,T=0)$ and
$\Xi\Omega-\Xi^*\Omega(S=1,T=1/2)$ systems 
are bounder than the $\Lambda\Lambda(S=0,T=0)$ and
$\Xi\Omega(S=1,T=1/2)$ systems due to the channel coupling effect.

\vspace{0.3cm}

In summary, we predict that $N \Omega(S=2,T=1/2)$, $\Delta\Omega(S=3,T=3/2)$,
$\Xi\Xi(S=0,T=1)$ and $\Xi\Omega(S=1,T=1/2)$ are weakly bound baryon-baryon
states. The mass of the $H$ particle is around the $\Lambda\Lambda$ threshold. 
$\Xi\Omega-\Xi^*\Omega(S=1,T=1/2)$ system is a bound state with
a relative large binding energy due to the strong channel coupling.

\vspace{0.3cm}

\noindent
{\bf 3.2. The systems in which the expectation values of $P^{sfc}_{36}$ operator
are close to $-1/9$.}

\vspace{0.3cm}

The systems in which $\langle P^{sfc}_{36} \rangle \approx -1/9$ are 
collected in this category. The binding energies, $E_b$, and corresponding 
root-mean-square radii, ${\cal R}$, of these systems are tabulated in
Table 3.  

\vspace{0.3cm}

\begin{scriptsize}
{\flushleft{\bf {Table 3. Binding energy, $E_b$, and corresponding $RMS$,  
${\cal R}$, for the systems with $\langle P^{sfc}_{36} \rangle=-1/9$}}. 
The units for $E_b$ and RMS are in $MeV$ and $fm$, respectively.}

\begin{center}
\tabcolsep 3pt
\begin{tabular}{|c|c||c|c|c|}
\hline
${\cal{S}}$ & System & $\langle P^{sfc}_{36} \rangle$ & Model I (Set I) &
Model I (set II) \\
  &  &  & $E_b~~//{\cal R}$ & $E_b~~//{\cal R}$ \\ \hline
  0 & $\Delta\Delta(S=3,T=0)~(d^{*})$ & $-1/9$ & 22.2~//1.01 
     & 18.5~//1.05\\  \cline{2-5}
  & $\Delta\Delta (S=0,T=3)$ & $-1/9$ & 16.0~//1.10 
     & 13.5~//1.14\\  \hline
 -1 & $\Sigma^* \Delta (S=0,T=5/2)$ & $-1/9$
     & 24.6~//0.99 & 19.0~//1.04\\  \cline{2-5}
    & $\Sigma^* \Delta (S=3,T=1/2)$ & $-1/9$
     & 25.9~//0.95 & 29.3~//0.93\\  \hline
 -5 & $\Xi^* \Omega(S=0,T=1/2)$ & $-1/9$
     & 92.4~//0.71 & 76.5~//0.72 \\ \hline
 -6 & $\Omega\Omega(S=0,T=0)$ & $-1/9$
     & 116.1~//0.66 & 98.5~//0.67 \\ \hline
\end{tabular}
\end{center}
\end{scriptsize}
%
%\vspace{-0.5cm}
%
%{\flushleft {\scriptsize{* Units for $E_b$ and RMS are in $MeV$ and $fm$, 
%respectively.}}}

\vspace{0.5cm}

\noindent
>From this table, one sees that all the systems in this category have the
feature of $\langle P^{sfc}_{36}\rangle=-1/9$. It indicates that 
the system has relatively higher anti-symmetry in the $spin$-$flavor$-$color$
space. As a consequence, the contribution of the kinetic energy to the
binding energy plays a  relatively attractive role in comparision with
the kinetic energy in two independ baryons \cite{zyc} (see Appendix). This
characteristics would bring six quarks closer. If the chiral fields can
additionally provide attraction between interacting baryons, the deeply bound
state may be established, such as $\Omega\Omega(S=0,T=0)$ and $\Xi^*
\Omega(S=0,T=1/2)$. Even the contributions of chiral fields are mild, the
system with the symmetry  structure of this kind may still have the binding
energy of several tens  $MeV$. Therefore, in any case, the system in this
category would be a  bound state. However, it should be noticed that in most
states here, such as  $\Delta\Delta$ and $\Sigma^{*} \Delta$, both composed
baryons have strong  decay modes. Therefore, if the  $\Delta\Delta$ and
$\Sigma^{*} \Delta$ are not bound deeply enough, namely their masses are not
smaller than the thresholds of $N N \pi \pi$ and $N \Lambda \pi \pi$,
respectively, these four states should be bound states with broad widths.
Only $\Omega\Omega(S=0,T=0)$ and $\Xi^* \Omega(S=0,T=1/2)$ are the most
interesting systems. Both $\Omega$'s in  $\Omega\Omega(S=0,T=0)$ can decay
only through the weak mode, so that $\Omega\Omega(S=0,T=0)$ is a bound state
with a narrow width. It is also possible that the mass of  $\Xi^*
\Omega(S=0,T=1/2)$ is smaller than the threshold of $\Xi \Omega \pi$ (with
Set I), then this state could also be a narrow width bound state. 
\vspace{0.5cm} 

\noindent
{\bf 4. Effects of interactions induced by chiral-quark field couplings.}

\vspace{0.3cm} 

In this section, we would demonstrate another factor which dominates
the binding behavior of the system concerned, namely the interactions 
caused by the chiral-quark field couplings.

\vspace{0.3cm}

The characters of chiral fiels are different case by case, and the
importance and sensitivity of these fields in the bound-state problem
of six-quark systems also dissimilar. In order to analyze the effect of the
chiral field on the binding energy of the baryon-baryon system, we build up 
other two models. In Model II, the $\pi$, $K$, $\eta$, $\eta^{\prime}$ and 
$\sigma_0$ are considered. This is so called extended
chiral $SU(2)$ quark model. The Model III is called chiral $SU(2)$ quark model,
in which only $\pi$ and $\sigma_0$ fields are remained. 

\vspace{0.3cm}

We arrange all concerned systems which are possibly bound into following
types: 
\vspace{0.3cm}

\noindent
{\bf 4.1. Deuteron-like systems.}

   To convince readers the the importance of the interaction for binding, we
present the contributions of various terms, such as kinetic energy, OGE,
pseudoscalar meson exchange and scalar mesons exchange terms to the total
binding energy for the typical case $\Xi\Xi_{(0,1)}$ in table 4. 

\begin{scriptsize} 
{\flushleft \bf Table 4. Contributions of various terms to binding energy 
for $\Xi\Xi_{(0,1)}$, the unit for energy is in $MeV$.} 
\begin{center} 
\begin{tabular}{rrrrrrrrrrr} \hline 
$~~~~~~~$      & kine.& OGE & $\pi$ & K & $\eta$ & $\eta'$ &
$\sigma$ & $\sigma'$ & $\kappa$ & $\epsilon$ \ \\ \hline   
Model I Set 1  &-17.3&-15.2&-1.2&-1.7&-1.4&-0.8&37.7&1.9&-1.2&6.2 \\ 
Model~~~~~~~~II&-12.0&-5.5&-0.8&-1.1&-0.8&-0.5&26.0&$-$&$-$&$-$\\  
Model~~~~~~~III&-16.1&-15.8&-1.4& $-$&$-$&$-$ &40.5&$-$&$-$&$-$\\ \hline
\end{tabular}
\end{center} 
\end{scriptsize} 

\vspace{0.3cm}

  From the table, we sees that although the kinetic energy, OGE and
pseudoscalar mesons provide the repulsive feature, the strong attractive
effect from scalar mesons, especially $\sigma$ meson, would compensate the
repulsion and make the $\Xi\Xi_{(0,1)}$ system weakly bound.

\vspace{0.3cm}

We then demonstrate the binding energies and the corresponding 
root-mean-square-radii of the bound systems in Sect.3.1 with
Models I (set I), II and III, respectively, in Table 5, except 
the $H$ particle which was carefully studied in our previous 
paper \cite{szy}.

\vspace{0.3cm}

\begin{scriptsize}
{\flushleft{\bf {Table 5. Binding energy, $E_b$, and corresponding $RMS$,  
${\cal R}$, for deuteron-like systems in various chiral quark models}}.
The units for $E_b$ and RMS are in $MeV$ and $fm$, respectively. 
The system is denoted by symbol $[B_1 B_2]_{(S,T)}$ with $B_1,~ B_2,~S$
and $T$ being baryons 1 and 2 and the total spin and isospin
of the system, respectively.}
\begin{center}
\tabcolsep 3pt
\begin{tabular}{|c||c|c|c|c|c|}
\hline
{$~$} &  ${\cal{S}}=-5$ &  ${\cal{S}}=-4$
& \multicolumn{2}{|c|}{${\cal{S}}=-3$}& ${\cal{S}}=0$
        \\ \cline{2-6}
Model & $[\Xi\Omega - \Xi^{*}\Omega]_{(1,1/2)}$ & $[\Xi\Xi]_{(0,1)}$ 
       & $[N \Omega]_{(2,1/2)}$ & $[\Delta\Omega]_{(3,3/2)}$ 
       & $d_{(S=1,T=0)}$\\
       & $E_b~~//{\cal R}$ & $E_b~~//{\cal R}$ & $E_b~~//{\cal R}$ 
       & $E_b~~//{\cal R}$ & $E_b~~//{\cal R}$ \\ \hline
I (Set I) & 32.9~//0.78 & 4.1~//1.17 &  3.5~//1.18 & 4.4~//1.15 
          & 0.2~//1.63\\ \hline
II        & 29.3~//0.79 & -0.5~//1.33 &  31.8~//0.81 & 34.3~//0.80 
          & 2.1~//1.52\\ \hline
III       & 17.6~//0.86 & 3.1~//1.18 &  49.5~//0.74 & 49.6~//0.74 
          & 4.4~//1.41\\ \hline
\end{tabular}
\end{center}
\end{scriptsize}

\vspace{0.5cm}

As mentioned in the preceding section, the binding energies of the systems
in this type are sensitive to the contributions offered by the chiral field.
This character is very similar to that of deuteron and can explicitly be seen in this table. 
In most cases in this type, say the systems with 
low strange number, when one switches the model
from Model II to Model I (Set I) or even from Model III to model II, 
namely the strange clouds are taken into account step by step, additional 
strange-cloud-caused 
interactions would make the overall inter-baryon interaction
less attractive in nature,  and consequently the systems would 
become less bounder.
In a word, these systems are bounder in the chiral $SU(2)$ quark model
than in the chiral $SU(3)$ quark model. But, in high strangeness systems,
additionally taking strange clouds into account would benefit the binding.

\vspace{0.3cm}

The channel sketches and the predicted mass ranges of these states with 
the different combinations of chiral fields, which are denoted by shaded areas,
are plotted in Figs.2(a)-(d) for the $[\Xi\Omega - \Xi^{*}\Omega]_{(1,1/2)}$,
$[\Xi\Xi]_{(0,1)}$, $[N \Omega]_{(2,1/2)}$ and $[\Delta\Omega]_{(3,3/2)}$
systems, respectively.

\vspace{0.5cm} 
%{\scriptsize {\bf Fig.2. The channel sketches and the predicted ranges 
%of states.} The shaded area denotes the predicted range of the 
%possible bound state.} 

\vspace{0.5cm}

\noindent
{\bf 4.2. $\Delta\Delta$- and $\Omega\Omega$-like systems}

   Now, we analyze the role or impoatance of the interaction in the system in
the third class or in subsection 3.2.. As is mentioned above, except the
symmtry character $\langle P_{36}^{sfc}\rangle\approx-1/9$ of the system,
the effect of the interaction is dominantly responsible for binding. A
detailed study of the binding energies of the systems in this category show
that, different with the contribution provided by the interaction in the
second class, the contribution from exchange term of the interaction is
substantially large and gernerally has two different character for various
states in the class. Considering this difference together with the different
decay mode, we further distinguish the states in this class into
$\Delta\Delta$-like states and $\Omega\Omega$-like states.
In $\Delta\Delta$-like states, where only weak decays exist, the
direct and exchange terms of interactions play comparable roles in forming
dibaryon and the binding energies of these states are only a few tens of MeV.
As a specific example, the calculated result showes that in
$\Delta\Delta_{(0,3)}$, the exchange term of the interaction induced by
$\sigma$ meson provides a binding energy of 17.0 MeV, which is comparable to
the direct contribution of 16.0 MeV. While in
$\Omega\Omega$-like states, where at most one strong decay mode exists, the 
exchange terms of interactions play much more important roles than the direct
terms do and the binding energies of these states can reach
nearly one hundred MeV. For instance, in $\Omega\Omega_{(0,0)}$, the
contribution of the exchange term of the $\sigma$ interaction is 80.6MeV,
which is nearly two times larger than that of the direct contribution of
44.3 MeV. We would investigate  these two types of states in detail
in the following. 

\vspace{0.5cm} 

\noindent

{\bf 4.2.a. $\Delta\Delta$-like systems.}

We firstly single out $\Delta\Delta_{(0,3)}$ as a representative to
decompose the binding energy of the state in this
class. The results are tabulated in Table 6.  

\begin{scriptsize} 
{\flushleft \bf Table 6. Contributions of various terms to binding energy 
for $\Delta\Delta_{(0,3)}$, the unit for energy is in $MeV$}. 
\begin{center} 
\begin{tabular}{rrrrrrrrrrr} \hline 
$~~~~~~~$      & kine.&  OGE & $\pi$ & K & $\eta$ & $\eta'$ &
$\sigma$ & $\sigma'$ & $\kappa$ & $\epsilon$ \ \\ \hline   
Model I Set 1  &8.0&-26.7&-15.0&0.0&-2.3&-0.4&33.0&14.5&0.0&4.8 \\ 
Model~~~~~~~~II&6.2&-12.7&-10.0&0.0&-1.2&-0.2&24.1&$-$&$-$&$-$\\  
Model~~~~~~~III&12.5&-17.8&-17.8& $-$&$-$&$-$ &36.4&$-$&$-$&$-$\\ \hline
\end{tabular}
\end{center} 
\end{scriptsize} 

\vspace{0.3cm}

   It is seen that only OGE and pseudosclar mesons provide repulsive
contribution, while both kinetic energy and the scalar mesons devote their
attractive contribution to binding. Moreover, it is shown that again the
potential terms, especially the terms provided by the scalar meson, dominate
the binding behavior.

\vspace{0.3cm}

The binding energies and the corresponding 
root-mean-square-radii for this type of states with
Models I (set I), II and III are tabulated in Table 7, respectively.

\vspace{0.3cm}

\begin{scriptsize}
{\flushleft{\bf {Table 7. Binding energy, $E_b$, and corresponding $RMS$,  
${\cal R}$, for $\Delta\Delta$-like systems in various chiral quark models}}.
The units for $E_b$ and RMS are in $MeV$ and $fm$, respectively.
The system is denoted by symbol $[B_1 B_2]_{(S,T)}$ with $B_1,~ B_2,~S$
and $T$ being baryons 1 and 2 and the total spin and isospin
of the system, respectively.}

\begin{center}
\tabcolsep 3pt
\begin{tabular}{|c||c|c|c|c|}
\hline
{$~$}  & \multicolumn{2}{|c|}{${\cal{S}}=-1$}
& \multicolumn{2}{|c|}{${\cal{S}}=0$}\\
         \cline{2-5}
 Model & $[\Sigma^{*} \Delta]_{(0,5/2)}$ & $[\Sigma^{*} \Delta]_{(3,1/2)}$ 
       & $[\Delta\Delta]_{(0,3)}$ & $[\Delta\Delta]_{(3,0)}~~~(d^{*})$ \\ 
       & $E_b~~//{\cal R}$ & $E_b~~//{\cal R}$ & $E_b~~//{\cal R}$ 
       & $E_b~~//{\cal R}$ \\ \hline
I (Set I) & 24.6~//0.99 & 25.9~//0.95 & 16.0~//1.10 & 22.2~//1.01 \\ \hline
II        & 11.7~//1.13 & 78.0~//0.79 &  6.3~//1.25 & 64.8~//0.84 \\ \hline
III       & 21.5~//0.99 & 76.3~//0.80 & 13.2~//1.11 & 62.7~//0.86 \\ \hline
\end{tabular}
\end{center}
\end{scriptsize}

\vspace{0.5cm}

It is shown that the binding characters vary with the spin
structures of systems. When one turns the model from Model I (Set I) to 
Model II, the binding energies of the $S=0$ systems become smaller and
those of the $S=3$ systems become larger. However, if the model is switched
from Model II to Model III, in comparison with the preceding
case the binding energies for different spin systems change
in opposite directions. Nevertheless, substituting Model I (Set I) by Model III
would cause the binding energy decreasing in the $S=0$ system and 
increasing in the $S=3$ system. Precisely, the systems with $S=0 $ and $S=3$
are more bound in the chiral $SU(3)$ quark model and the extended chiral
$SU(2)$ quark model, respectively. Moreover, because $CMI$ shows repulsive
feature in the $S=0$ system and attractive nature in the $S=3$ system, the 
binding energies of the $S=3$ systems are generally higher than those of 
the $S=0$ systems.

\vspace{0.3cm}

We also draw channel sketches and predicted mass ranges of
 $[\Sigma^{*} \Delta]_{(0,5/2)}$, $[\Sigma^{*} \Delta]_{(3,1/2)}$,
 $[\Delta\Delta]_{(0,3)}$, $[\Delta\Delta]_{(3,0)}$ states with different
 combination of chiral fields, which are signified by shaded areas, 
in Fig.2(e) and Fig.2(f), respectively.
 Since in the 
$\Sigma^{*}\Delta$ cases, the predicted masses are 
above the threshold of the strong decay channel $\Lambda N \pi\pi$, and in 
the $\Delta\Delta$ cases, the resultant masses are above
the threshold of the strong decay channel $N N \pi\pi$, these bound states
would have large widths. It might not be easy to detect in experiments.

\vspace{0.3cm}

It is worth to mention that because the Hamiltonian for the octet and decuplet
baryons are in the same form and the calculated masses of these baryons can
well-fit the data, the RGM calculations for the systems involving $\Delta$ are
valid. Of course, whether this model treatment agrees with reality should be
verified by the experiment in future.

\vspace{0.5cm}

\noindent
{\bf 4.2.b. $\Omega\Omega$-like systems.}

    As what we did for deuteron- and $\Delta\Delta$-like states, we
choose $\Omega\Omega_{(0,0)}$ as a specific example to show the 
contributions of various Hamiltonian terms.  

\begin{scriptsize} 
{\flushleft \bf Table 8. Contributions of various terms to binding energy 
for $\Omega\Omega_{(0,0)}$, the unit for energy is in $MeV$}. 
\begin{center} 
\begin{tabular}{rrrrrrrrrrr} \hline 
$~~~~~~~$      & kine.& OGE & $\pi$ & K & $\eta$ & $\eta'$ &
$\sigma$ & $\sigma'$ & $\kappa$ & $\epsilon$ \ \\ \hline   
Model I Set 1  &8.6&-82.7&0.0&0.0&-7.6&-15.0&124.9&0.0&0.0&87.7 \\ 
Model~~~~~~~~II&12.3&-33.3&0.0&0.0&-6.5&-12.5&112.5&$-$&$-$&$-$\\  
Model~~~~~~~III&12.7&-58.5&0.0&$-$&$-$&$-$ &97.7 &$-$&$-$&$-$\\
\hline \end{tabular}
\end{center} 
\end{scriptsize}

\vspace{0.3cm} 

The large binding energy of $\Omega\Omega_{(0,0)}$ in model I and
the binding behaviors of the system in model II and III can clearly be
undestood by  the detailed contributions of interactions in table 8.
Apparently, the kinetic energy contribute attractively to the binding energy
and the $\sigma$ meson (in Model I , also $\epsilon$ meson) provides strong
attractive contribution and dominates the binding behavior. OGE, $\eta$ and
$\eta'$ are repulsive behavior contributors.

\vspace{0.3cm}

We tabulate the binding energies and the corresponding 
root-mean-square-radii for this type of systems with
Models I (set I), II and III in Table 9, respectively.

\vspace{0.3cm}

\begin{scriptsize}
{\flushleft{\bf {Table 9. Binding energy, $E_b$, and corresponding $RMS$,  
${\cal R}, $ for $\Omega\Omega$-like systems in various chiral quark models}}.
The units for $E_b$ and RMS are in $MeV$ and $fm$, respectively.
The system is denoted by symbol $[B_1 B_2]_{(S,T)}$ with $B_1,~ B_2,~S$
and $T$ being baryons 1 and 2 and the total spin and isospin
of the system, respectively.}

\begin{center}
\tabcolsep 3pt
\begin{tabular}{|c||c|c|}
\hline
{$~$}  & {${\cal{S}}=-6$} & {${\cal{S}}=-5$} \\ \cline{2-3}
 Model & $[\Omega\Omega]_{(0,0)}$ & $[\Xi^{*}\Omega]_{(0,1/2)}$ \\
       & $E_b~~//{\cal R}$ & $E_b~~//{\cal R}$ \\ \hline
I (Set I) & 116.1~//0.66 & 92.4~//0.71 \\ \hline
II        &  74.2~//0.69 & 58.8~//0.75 \\ \hline
III       &  53.5~//0.73 & 52.1~//0.76 \\ \hline
\end{tabular}
\end{center}
\end{scriptsize}

\vspace{0.5cm}

It is seen that no matter in which model, these two states are bound states, or 
dibaryons. The $CMI$ of the system in this type shows 
repulsive feature which offers disadvantage to binding. However, 
the benefit from the symmetry structure discussed in Sect.3.2 somehow 
overcomes this disadvantage. On the other hand, because these systems are
high-strangeness systems, the attractive characteristics from the $\sigma_0$-
field-caused interaction and strange-cloud-induced interactions make the
systems deeply bound and corresponding ${\cal R}$s less than
$0.76fm$. In other word, by using the chiral $SU(3)$ quark model, these
systems could be even bounder. More detailed analysis for
$[\Omega\Omega]_{(0,0)}$ is given in Ref.[21].

\vspace{0.3cm}

The channel sketches and the predicted masses of  
the $[\Omega\Omega]_{(0,0)}$ and $[\Xi^{*}\Omega]_{(0,1/2)}$ states 
with different combinations of chiral fields, which are presented 
through shaded areas, are shown 
in Fig.2(g) and Fig.2(h), respectively. Since $\Omega$ does not
have strong decay mode, the predicted
 $[\Omega\Omega]_{(0,0)}$ bound state should have narrow width.
However, in the $[\Xi^{*}\Omega]_{(0,1/2)}$ system, because
$\Xi^{*}$ has the strong decay mode and the predicted masses of 
$\Xi^{*}\Omega$ by using different models stride over the threshold of the 
strong decay channel 
$\Xi\Omega\pi$, whether the width of the bound state is narrow
depends on the model employed. With the chiral
$SU(3)$ model, which, we believe, is a right model for the high strangeness system,
the width should be narrow too.

\vspace{0.3cm}

Furthermore, we would emphasize that the purpose of this section is to
trace the influences of various chiral fields on the calculated results through 
employing different models such as 
Models II and III. In our consideration, the result with Model I is the
most reliable one, because investigating a system with strange quarks
without considering strange clouds is unreasonable.

\vspace{0.5cm}

\noindent
{\bf 5. Model parameter dependence}

To ensure the reliability of our predictions, various kinds of model 
parameter-dependencies are also investigated.
%Finally, the influences on predicted masses in the chiral $SU(3)$ quark model,
%which are induced by the variations of input
%parameters $b_u$ and $m_s$ are explored.

\vspace{0.3cm}

It is clear that the variation of $b_u$ would change the size of the baryon
as well as the values of model parameters via the conditions mentioned in
Sect.2, and the variation of $m_s$ would change the strength of the
OGE potential, $g_s$. We plot $E_b$ with respect to $b_u$
with different $m_s$ for six most interesting systems, where 
$\langle P_{36}\rangle=-1/9$, in Fig.3. From this 
figure, one finds that the binding energy decreases with increasing 
values of $b_u$. When $b_u$ is altered to a large value, the higher spin the
system has, the milder the binding energy changes, while with increasing the 
mass of strange quark,
the more strange quarks the system contains, the larger binding energy 
the system has. As long as the value stays in a reasonable region, 
say from $0.5fm$ to $0.55fm$, these systems remain the bound nature. 

\vspace{0.3cm}

The $m_{\sigma}$-dependence is also studied, because the $\sigma$ induced 
interaction is always a dominant piece that responses for the binding 
behavior for the two-baryon system. The results with different values of
$m_{\sigma}$ are tabulated in Table 7.

\vspace{0.5cm}

\begin{scriptsize}
{\flushleft {\bf {Table 10. The $m_{\sigma}-$dependence of the binding energy 
$E_b$. }}  The units of $E_b$ and $m_{\sigma}$ are in $MeV$.
The system is denoted by symbol $[B_1 B_2]_{(S,T)}$ with $B_1,~ B_2,~S$
and $T$ being baryons 1 and 2 and the total spin and isospin
of the system, respectively.}

\begin{center}
\tabcolsep 3pt
\begin{tabular}{|c|c||c|c|c|c|c|}
\hline
\multicolumn{2}{|c||}{$m_{\sigma}$} & 550 & 600 & 625 & 650 & 700\\ \hline
    & ${\Omega\Omega_{(0,0)}}$     & 140.6 & 121.9 & 116.1 & 107.1 & 94.3\\
   \cline{2-7}
    & ${\Xi^{*}\Omega_{(0,1/2)}}$  & 116.2 & 99.5  & 92.4  & 86.5  & 75.4\\
   \cline{2-7}
    & ${\Xi\Omega_{(1,1/2)}}$      &  21.3 & 11.2  &  9.5  &  4.7  &  0.5\\
   \cline{2-7}
    & ${N \Omega_{(2,1/2)}}$     & 17.7  &  7.1  &  3.5  &  0.9  & -2.8\\
   \cline{2-7}
$E_b$ & $\Sigma^{*}\Delta_{(0,5/2)}$ & 36.5  & 28.1  & 24.6  & 21.9  & 17.0 \\
    \cline{2-7}
   & $\Sigma^{*}\Delta_{(3,1/2)}$ & 40.8  & 30.2 & 25.9  & 22.5  & 16.4 \\
    \cline{2-7}
   & ${\Delta\Delta_{(0,3)}}$     & 25.5  & 18.7  & 16.0  & 13.9  & 10.2\\
    \cline{2-7}
   & ${\Delta\Delta_{(3,0)}}$     & 35.0  & 25.8  & 22.2  &  19.3 & 14.2\\
   \cline{2-7}
    & ${NN_{(1,0)}}~(d)$          & 5.1   & 1.6   &  0.2  & -0.5  & -2.1\\
   \hline
\end{tabular}
\end{center}
\end{scriptsize}

\vspace{0.3cm}

>From this table, one sees that as long as $m_{\sigma}$ is in a reasonable 
region, namely the existent experimental data can unifiably and reasonably
be explained, most systems studied would not change their binding behaviors,
except those whose binding energies are marginal, for instance, when
$m_{\sigma}=700MeV$, $N \Omega$ becomes unbound.

\vspace{0.3cm}

The cutoff mass dependence is also investigated. Our result shows that if one 
increases all the cutoff masses to $1.2GeV$, the binding energies of the 
systems concerned would slightly arisen. The binding behaviors are relative 
stable.

\vspace{0.3cm}

In a word, as long as the model parameters are
selected in reasonable regions and all the experimental data, such as $NN$
and $NY$ scattering data, the ground state masses of baryons, binding
features of well-known baryon-baryon states, and etc., can be basically 
reproduced, the concerned six two-baryon systems with
$\langle P^{sfc}_{36} \rangle=-1/9$ should be bound, especially, 
$\Omega\Omega_{(0,0)}$ and $\Xi^{*}\Omega_{(0,1/2)}$ should be deeply bound
dibaryons.

\vspace{0.3cm}

%{\scriptsize {\bf Fig.3. $b_{u}-$ and $m_{s}-$dependencies of $E_b$ for 
%six most interesting systems.}}

\vspace{0.5cm}

\noindent
{\bf 6. Concluding remarks.}

\vspace{0.3cm}

The possible candidates of $S-$wave baryon-baryon bound states with 
various strange numbers
are systematically studied by using the chiral quark model. 
In terms of the chiral $SU(3)$ quark model, which are believed to be
one of the best models in the quark-gluon degrees of freedom in well 
reproducing as much experimental data as possible, we predict the two-baryon 
bound states without any additional parameters. 

\vspace{0.3cm}

There are two major factors that affect the binding behavior of the two-baryon
system. One is the symmetry property of the system, which is characterized by
the matrix-element $\langle {\cal A}^{sfc}_{36} \rangle$, and the other is the 
interactions between quarks including both the direct and exchange components,
especially the interactions caused by chiral-quark field couplings. 

\vspace{0.3cm}  

We classify two-baryon bound states (even almost bound states) into three
types. The first type state is the deuteron-like state. The symmetry structure
of the system in this type shows that the quark exchange effect is not so 
important, namely, there is no serious Pauli blocking effect. If the chiral
field can provide enough attraction between interacting baryons, 
the bound state can be formed. This kind of bound state might also be obtained
by using the models on baryon level. For instance, in terms of Nijmegen model,
one also predicted some bound states in the ${\cal{S}}=-2,~-3$ and $-4$
systems \cite{rk}.

\vspace{0.3cm}

The $\Delta\Delta$-like state is the second type state. The system of this 
type is mostly symmetric in the $orbit$ space. In these states, the interaction, 
especially the scalar meson induced interaction, dominates the binding behavior. 
Together with the symmetry behavior of the system, the relatively deeply-bound 
state can be formed. However, because
both baryons in the system have strong decay modes, only when the predicted
binding energy is lower than the threshold of the strong decay channel, say the 
$N N \pi\pi$ channel for the $\Delta\Delta$ system and the $\Lambda N \pi\pi$ 
channel for the $\Sigma^{*}\Delta$ system, the width of the bound state
could be narrow. In our calculation, the predicted binding energies
of $\Delta\Delta$ and $\Sigma^{*} \Delta$ are not large enough, so that the 
widths of these bound states should be rather broad. Although this kind 
of bound state might not easily be detected in experiments, it may worth
to search in the future.

\vspace{0.3cm}

The third type state is the $\Omega\Omega$-like state, which is the most 
interesting state in our study. Same as those in the second type,
the system in this type is mostly symmetric in the $orbit$ space.
Meanwhile, the strange chiral fields can offer rather strong attraction.
As a result, these states are deeply bound states. If we believe that
the chiral $SU(3)$ quark model is one of the most suitable models in describing
the system with high strangeness, the predicted binding energy of 
$\Xi^{*} \Omega$ is possibly below the threshold of the strong decay 
channel $\Xi \Omega \pi$. Thus, both $\Omega\Omega$ and $\Xi^{*} \Omega$
can merely have weak decays, and consequently are deeply bound states 
with narrow widths.

\vspace{0.3cm}

It should be specially emphasized that the states of the second and third 
types possess the six-quark structure, and their inter-baryon distances
are relatively short. These characteristics cannot be provided by the model
on the baryon level. Thus, they are new dibaryon systems. The existences of
this kind of dibaryons would be an important place to reveal $QCD$ 
phenomenology.

\vspace{1.0cm}

\noindent
{\bf Appendix.}

\vspace{0.5cm}

In the framework of RGM \cite{tang}, the upper bound is given by the expectation 
value of the Hamiltonian
\begin{eqnarray}
\langle H \rangle =\frac{\langle \Psi | H |\Psi \rangle }{\langle \Psi |
\Psi \rangle },
\end{eqnarray}
where $H=T+V$ is the Hamiltonian, with $T$ and $V$ being the kinetic
energy operator and the potential operator, respectively, $\Psi$ represents
the trial wave function of  the system and $\langle H \rangle $ denotes the
upper bound of the system. The kinetic operator of a six-quark system can be
written as:
\begin{eqnarray}
T=\sum_{i=1}^{6} T_{i} - T_{CM},
\end{eqnarray}
where $T_{i}$ and $T_{CM}$ denote the kinetic energy operators of the i-th
quark and of the center of mass motion (CM), respectively. 
Substituting the trial wave function (Eq.(21)) and the antisymmetrizer
${\cal A^{\prime}}$ (Eq.(22)) into projection equation
\begin{eqnarray}
\langle
[ \hat{\phi}_{A} \hat{\phi}_{B}
\chi_{CM}(\vec{R_{CM}})]_{ST} |H-E| {\cal A}^{\prime}
[ \hat{\phi}_{A} \hat{\phi}_{B}
\chi_{rel}(\vec{R})\chi_{CM}(\vec{R_{CM}})]_{ST} \rangle = 0, 
\end{eqnarray}
one obtains RGM equation
\begin{eqnarray}
\big\{ &-& \frac{\hbar^2}{2\mu}\bigtriangledown^{2}_{\vec{R}}+
V^{dir}_{rel}(\vec{R})
 - E_{rel} \big\} \chi_{rel}(\vec{R})\\  \nonumber
 &+& \int [ K^{T}(\vec{R},\vec{R^{\prime}})
 + K^{V}(\vec{R},\vec{R^{\prime}}) - E_{tot}
    N^{exch}(\vec{R},\vec{R^{\prime}}) ]\chi_{rel}(\vec{R^{\prime}})
    ~d\vec{R^{\prime}}=0,
\end{eqnarray}
where $ V^{dir}_{rel}(\vec{R})$ represents the relative potential between
two clusters, $ E_{tot}$ and $ E_{rel}$ denote the total energy and the 
relative energy between two clusters, respectively, and 
$ K^{T}(\vec{R},\vec{R^{\prime}}),~ K^{V}(\vec{R},\vec{R^{\prime}})$ and
$ N^{exch}(\vec{R},\vec{R^{\prime}})$ describe the kinetic energy exchange
kernel, the potential energy exchange kernel and the normalization exchange
kernel, respectively. Expanding $\chi_{rel}(\vec{R})$ by well-defined basis
functions
\begin{eqnarray}
\chi_{rel}(\vec{R^{\prime}})=\sum_{i=1}^{n}c_{i}\phi_{rel}(\vec{R^{\prime}},
\vec{S_{i}}),
\end{eqnarray}
%with
%\begin{eqnarray}
%\phi_{rel}(\vec{R^{\prime}},\vec{S_{i}})=N_{rel}
%e^{-\frac{1}{2}\omega\mu_{AB}
%(\vec{R^{\prime}} - \vec{S_{i}})^2 },
%\end{eqnarray}
and left-multiplying $ \phi_{rel}(\vec{R},\vec{S_{i}})$
to RGM equation and integrating over
$\vec{R}$ and $\vec{R^{\prime}}$, one obtains the secular equation
of the bound state problem,
\begin{eqnarray}
&~&\sum_{i=1}^{n}[ H_{ij}-E_{tot}~N_{ij} ]~c_{j}\\ \nonumber
&=&\sum_{i=1}^{n}[ H^{\prime}_{ij}-E_{rel}~
N_{ij} ]~c_{j}=0,~~~~~(i=1,\cdots,n)
\end{eqnarray}
with $ H_{ij}$ and $ N_{ij}$ being the Hamiltonian and normalization kernels,
respectively, and
\begin{eqnarray}
E_{tot}=E_{in}+E_{rel},
\end{eqnarray}
$E_{tot}$, $E_{in}$ and $E_{rel}$ being the total, inner cluster and 
relative energies, respectively. Apparently, $ H^{\prime}$ can be written as:
\begin{eqnarray}
H^{\prime}_{ij}&=&H_{ij}-E_{in}N_{ij}\\ \nonumber
&=& [T_{ij}-E^{T}_{in}N_{ij}]+ [V_{ij}-E^{V}_{in}N_{ij}].
\end{eqnarray}
In this equation, $ T_{ij}$ and $V_{ij}$ is the kinetic energy and
potential energy kernels, respectively, and the superscripts $T$ and $V$
denote the kinetic energy and potential energy parts, respectively.
Then,
\begin{eqnarray}
H^{T\prime}_{ij}&=& T_{ij}-E^{T}_{in}N_{ij}\\ \nonumber
&=&(T^{dir}_{rel})_{ij}+(T^{exch}_{rel})_{ij}
\end{eqnarray}
with
\begin{eqnarray}
(T^{dir}_{rel})_{ij}=\langle \phi_{rel}(\vec{R^{\prime\prime}},\vec{S_{i}})|
(- \frac{\hbar^2}{2\mu}\bigtriangledown^{2}_{\vec{R^{\prime}}})
\delta(\vec{R^{\prime\prime}}-\vec{R^{\prime}})|
\phi_{rel}(\vec{R^{\prime}},\vec{S_{j}})\rangle
\end{eqnarray}
and
\begin{eqnarray}
(T^{exch}_{rel})_{ij}=\langle \phi_{rel}(\vec{R^{\prime\prime}},\vec{S_{i}})|
K^{T}(\vec{R^{\prime\prime}},\vec{R^{\prime}})|
\phi_{rel}(\vec{R^{\prime}},\vec{S_{j}})\rangle. 
\end{eqnarray}

\vspace{0.3cm}

To compute $H^{T\prime}_{ij}$ explicitly, we employ 
the Generating Coordinate Method (GCM) technique, which is equivalent
to RGM. In GCM, we re-write 
\begin{eqnarray}
\Psi=\sum_{i=1}^{n} c_{i} \psi_{i}
\end{eqnarray}
with
\begin{eqnarray}
\psi_{i}&=&{\cal A}^{\prime}[ \hat{\phi}_{A} \hat{\phi}_{B}
\phi_{rel}(\vec{R^{\prime}},\vec{S_{i}})\chi_{CM}(\vec{R_{CM}})
\chi_{ST}\chi_{c}]\\ \nonumber
&=&{\cal A}^{\prime}\big[ \prod_{n=1}^{3}\phi_{0s} \big( \vec{r}_{n}-
\frac{\vec{S}_{i}}{2}\big) 
\prod_{k=4}^{6}\phi_{0s} \big( \vec{r}_{k}+\frac{\vec{S}_{i}}{2}\big)
\chi_{ST}\chi_{c} \big]
\end{eqnarray}
and 
\begin{eqnarray}
\phi_{0s} \big( \vec{r}_{n}-\frac{\vec{S}_{i}}{2}\big)
=N_{n}\exp{-\frac{1}{2b_{n}^2}\big(\vec{r_{n}}-\frac{\vec{S}}{2}\big)^2}.
\end{eqnarray}
To ensure the total $CM$ motion can explicitly be separated, we keep
$\omega$ unchanged, namely 
\begin{eqnarray}
\omega=(m_{u}b_{u}^{2})^{-1}=(m_{s}b_{s}^{2})^{-1}.
\end{eqnarray}
If one temporarily ignores the flavor symmetry breaking, namely 
$b=b_{u}=b_{s}$ and $m=m_{u}=m_{s}$, for instance
in the $(\Omega\Omega)$ or $\Delta\Delta$ or $NN$ system, the following
discussion would be flavor independent and it is 
easy to obtain the kinetic energy
induced S-wave adiabatic {\it effective interaction} 

\begin{eqnarray}
E^{T}_{ii}&=& \frac{H^{T\prime}_{ii}}{N_{ii}}-\frac{K_0}{3}\\ \nonumber
&=& 2K_{0}x \big[ -1 + Ctgh(3x)
\frac{1-9\langle P^{sfc}_{36}\rangle Cosh(x)/Cosh(3x)}{1-
27\langle P^{sfc}_{36}\rangle Sinh(x)/Sinh(3x)} \big]
\end{eqnarray}
with
\begin{eqnarray}
x=\frac{S_{i}^{2}}{4b^{2}}
\end{eqnarray}
and
\begin{eqnarray}
K_{0}=\frac{3}{4}\hbar \omega.
\end{eqnarray}

\vspace{0.3cm}

    According to the definition of the binding energy
\begin{eqnarray}
E_b=-(M_{DB}-M_A-M_B),
\end{eqnarray}
where $M_{DB}$, $M_A$ and $M_B$ denote the masses of the
dibaryon, baryon A and baryon B, respectively, the contribution of the kinetic
energy to the binding energy of the dibaryon in the adiabatic approximation
should be
\begin{eqnarray}
E_b^T& = & -\sum_{i=1}^{n} c_i^{2}E^{T}_{b}(S_{i}) 
\end{eqnarray} 
with
\begin{eqnarray}
E^{T}_{b}(S_{i})& = & E_{ii}^T+\frac{K_0}{3},
\end{eqnarray} 
where $c_i$ is the wavefunction. We plot $E^{T}_{b}(S_{i})$ 
versus $S_{i}$ in Fig.4 (for convenience, we drop the subscript $i$ in 
$S_{i}$ in Fig.4).
In this figure, the solid and dashed curves denote the results with 
$\langle P_{36}^{sfc}\rangle =-1/9$ and $\langle P_{36}^{sfc}\rangle =-1/81$,
respectively. Apparently, when $\langle P_{36}^{sfc}\rangle=-1/81 $, 
$E^{T}_{b}(S_{i})$ has less attractive character than that in the 
$\langle P_{36}^{sfc}\rangle=-1/9 $ case. And when  
$\langle P_{36}^{sfc}\rangle~\geq~0$, it would be even less attractive.

Apparently, $E_b^T$ is always a negative
number (or repulsive for binding). This indicates that, when off-diagonal
matrix elements contribution can be neglected under adiabatic approximation,
$E_b^T$ always has a repulsive effect.

\vspace{0.3cm}

    However, when calculating $E_b^T$, one should also
take the off-diagonal matrix elements into account. Further calculation
shows that the contribution of the off-diagonal matrix elements
differs substantially in the system with
$\langle{P_{36}^{sfc}}\rangle=-1/9$ to that in the system with
$\langle{P_{36}^{sfc}}\rangle\approx 0$. In the former case, it is so large
that $E_b^T$ can be changed to a positive value (or attractive for binding).
For instance, in the  $\Omega\Omega_{(0,0)}$ dibaryon, the contribution of the
diagonal matrix elements is $-145.6$ MeV while the contribution of the
off-diagonal matrix elements is $154.2$ MeV,
and consequently $E_b^T=8.6$ MeV. On the other hand, in the latter case, this
contribution is not large enough to change the sign of $E_b^T$. For instance,
in the deuteron case, the 
contribution of the diagonal matrix elements is $-57.5$ MeV while the 
contribution of the off-diagonal matrix elements is $43.0$ MeV, and
consequently $E_b^T=-14.5 MeV$. The effect of the off-diagonal
matrix elements can also simply be seen by diagonalizing the
$-(E^T_{ij}+\frac{K_0}{3}),$ called $\tilde{E}^{T}_{ij}$, matrix. In the
$\langle{P_{36}^{sfc}}\rangle=-1/9$ case, some of diagonalized matrix
elements $\tilde{E}^{T}_{ii}$ can be positive. Because 
\begin{eqnarray}
E_b^T=\sum_{i}^{n} \tilde{c}_{I}^{2}\tilde{E}^{T}_{ii},
\end{eqnarray}
where $\tilde{E}^{T}_{ii}$  is the diagonalized matrix element, and $\tilde{c}_{i}$
is the wave function after the unitary transformation that diagonalizes the
$-(E^{T}_{ij}+\frac{K_0}{3})$ matrix, the resultant $E_b^T$ becomes positive. 
While for the system with  $\langle{P_{36}^{sfc}}\rangle\approx 0$, the
diagonalized matrix elements are always negative, consequently, $E_b^T$ always
keeps a negative value.

\vspace{0.3cm}

     The physics picture of the positive contribution (ot attractive
contribution) of the kinetic energy operator to the total binding energy may be
understood as the following. In the dibaryon with
$\langle{P_{36}^{sfc}}\rangle=-1/9$, such as $\Omega\Omega_{(0,0)}$, six
quarks interfere with each other very strongly due to the large quark
exchange effect, so that they cannot move as freely as they do in two
independent baryons. So the expectation value of kinetic energy operator in
such system (which is definitely positive) would be smaller than the
kinetic energy in two independent baryons.
While in system with
$\langle{P_{36}^{sfc}}\rangle\approx 0$, the quark exchange effect is so
weaker that the expectation value of kinetic energy operator is always larger
than that the kinetic energy in two baryons.

\vspace{1.0cm}

\begin{small}

\end{small}

\newpage
{\bf {Figure captions}}

\noindent
Fig.1. $\Lambda$-$P$ scattering.

\noindent
Fig.2. Energy level sketches for a.[$\Xi\Omega$-$\Xi^*\Omega]_{(1,1/2)}$, b.
 [$\Xi\Xi]_{(0,1)}$, c.[$N\Omega]_{(2,1/2)}$, d.[$\Delta\Omega]_{(3,3/2)}$,
 e.[$\Sigma^*\Delta]_{(0,5/2)}$ and [$\Sigma^*\Delta]_{(3,1/2)}$, 
 f.[$\Delta\Delta]_{(0,3)}$ and [$\Delta\Delta]_{(3,0)}$,
 g.[$\Omega\Omega]_{(0,0)}$, h.[$\Xi^*\Omega]_{(0,1/2)}$.

\noindent 
Fig.3. Binding energy $E_b$ with respect to $b_u$ and $m_s$ for systems with
 $\langle{\cal A}^{sfc}\rangle\approx$ 2. The solid curves are for $m_s$=470 MeV,
 the dashed curves are for $m_s$=515 MeV.

\noindent
Fig.4. In the S-wave case, the contribution to binding energy provided
by the kinetic-energy related terms with different $S$ in the adiabatic 
approximation.

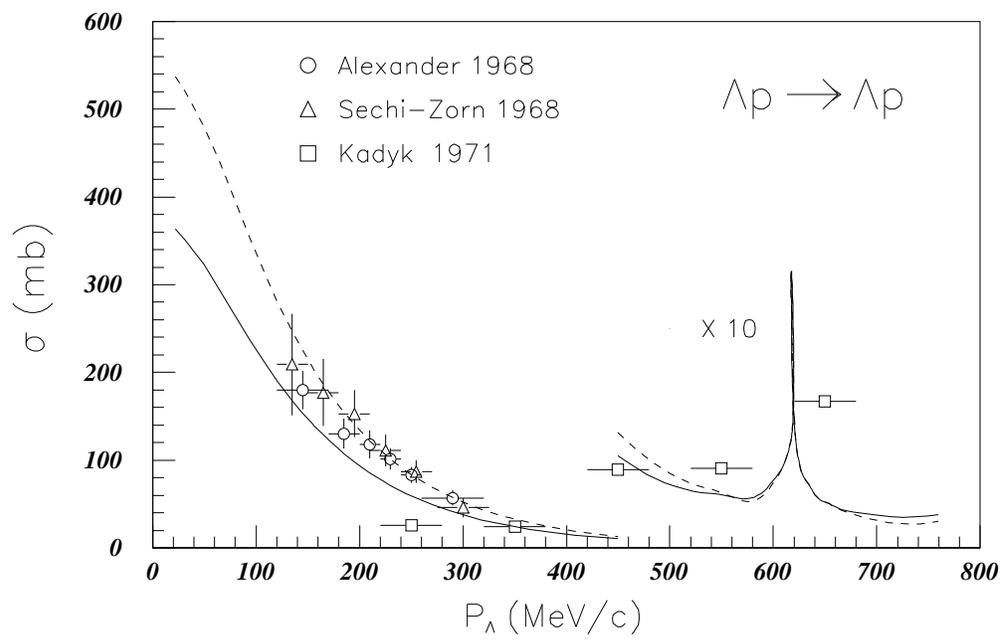
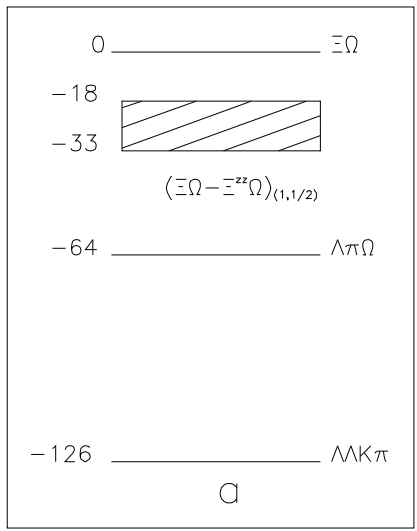
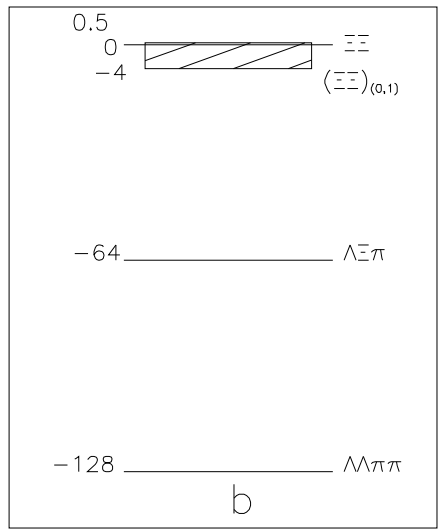
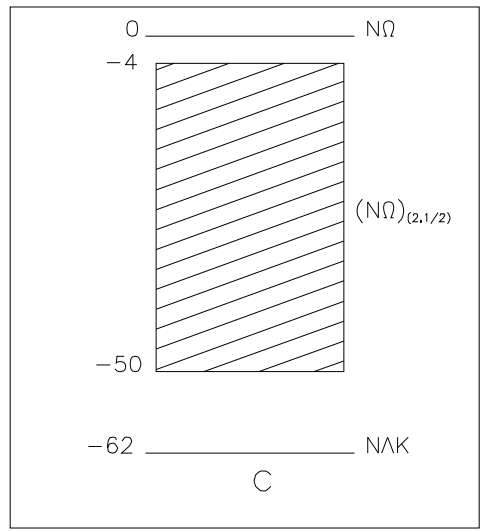
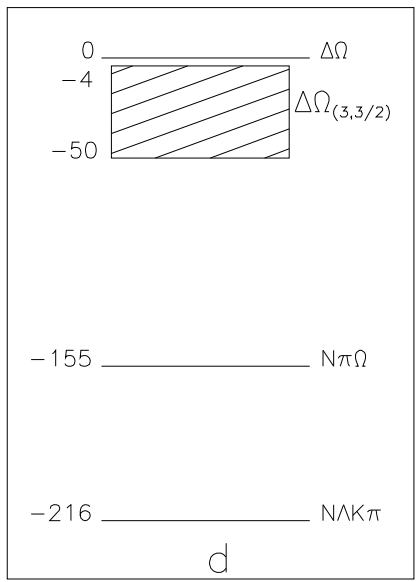
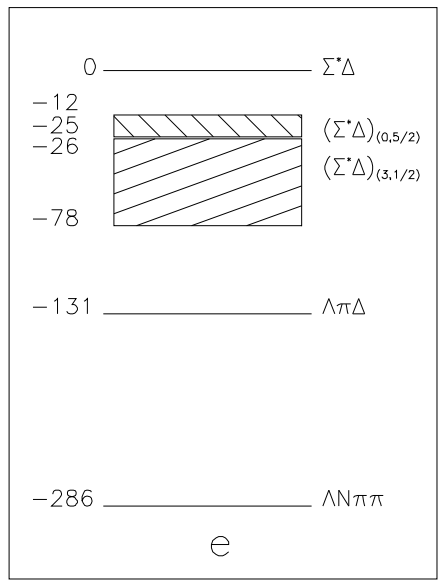
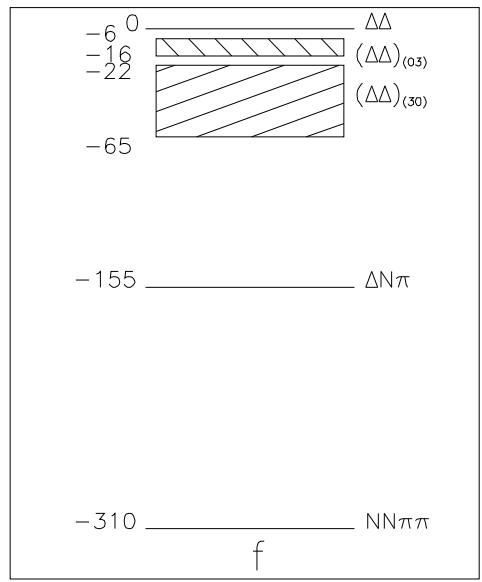
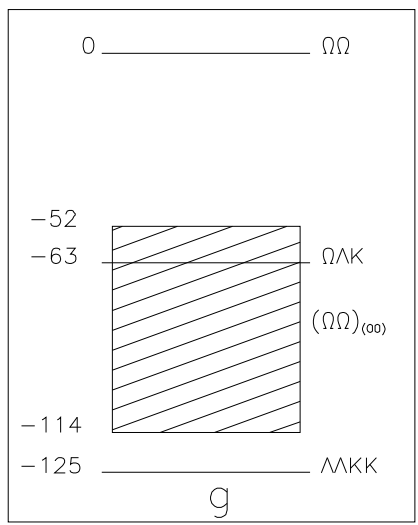
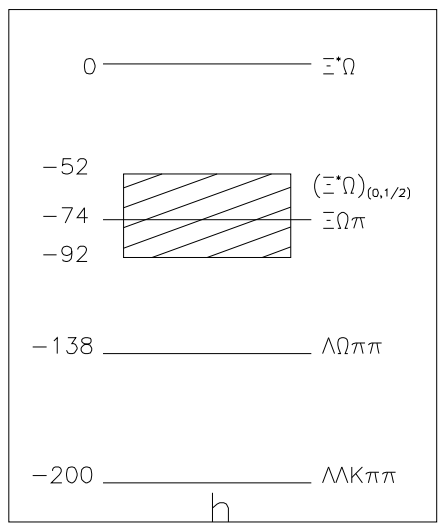
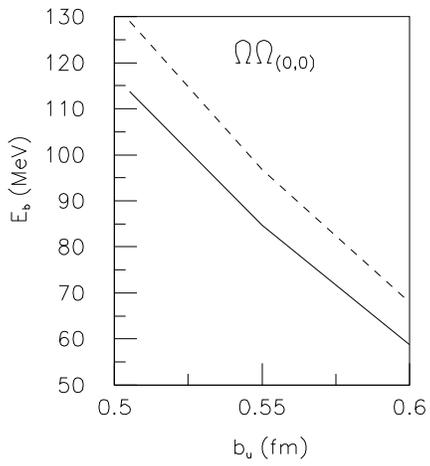
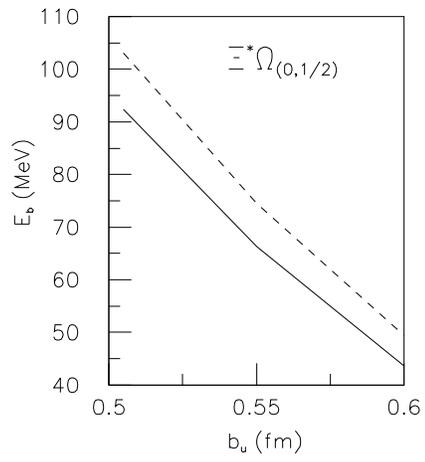
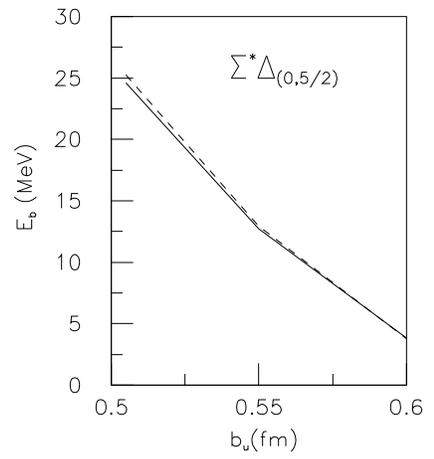
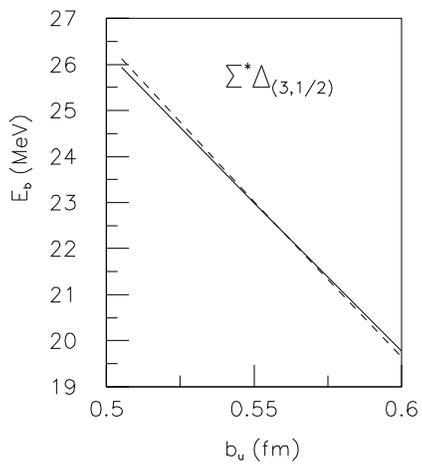
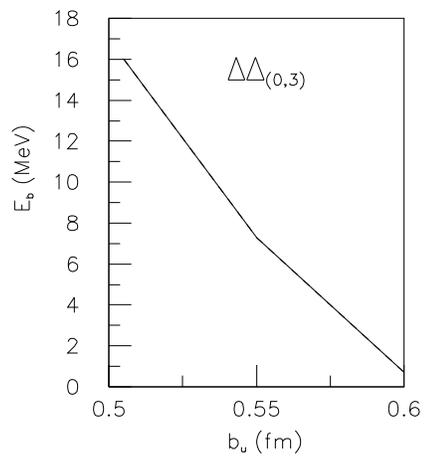
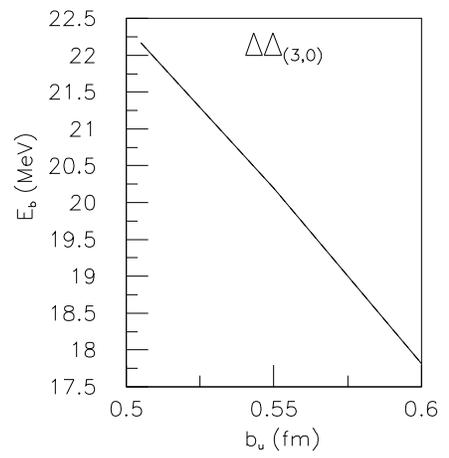
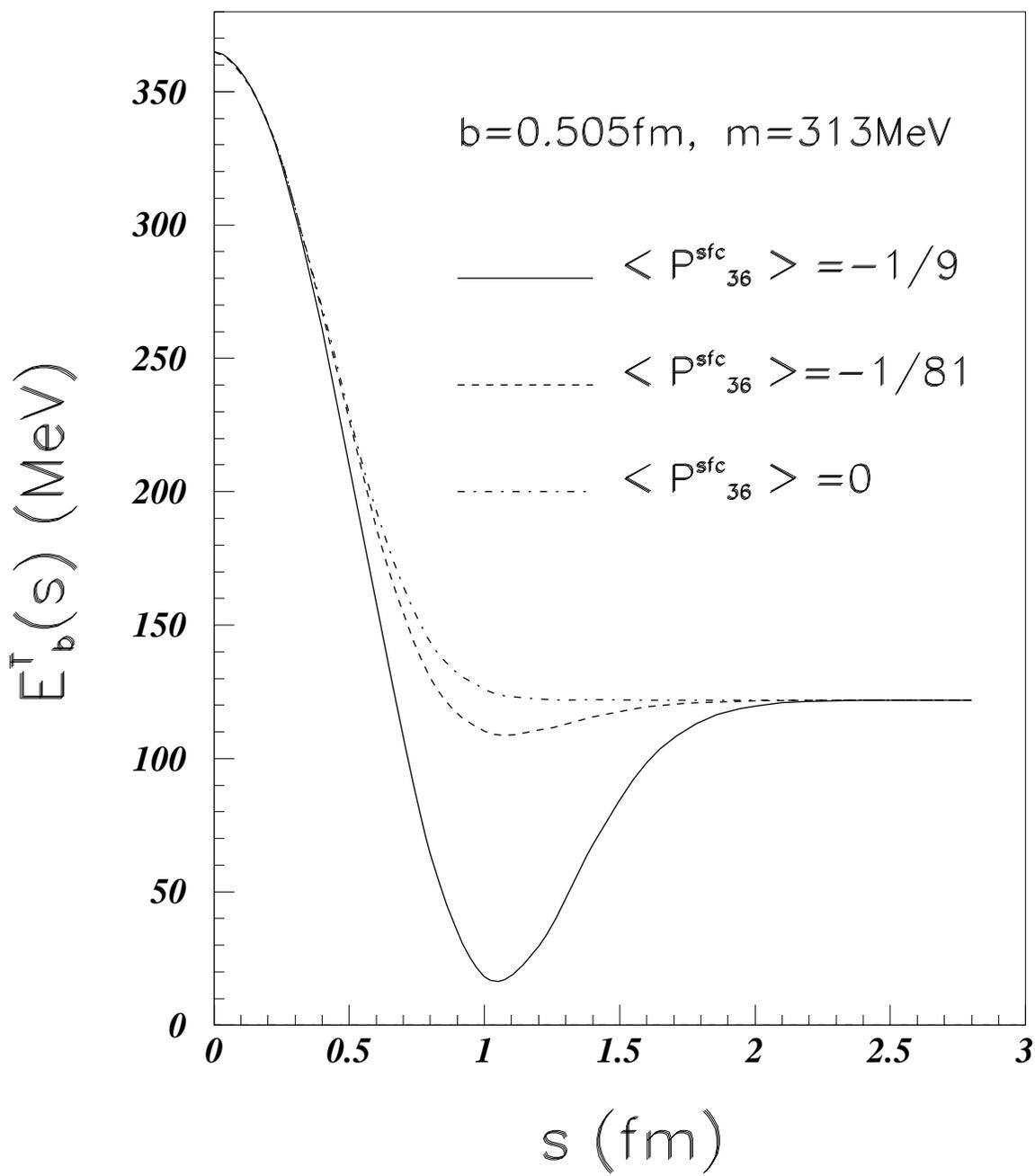
\end{document}